\documentclass[letterpaper,twocolumn,10pt]{article}
\usepackage{usenix,epsfig,verbatim}
\usepackage{titlesec}
\usepackage{xspace}
\usepackage{paralist}

\usepackage{xcolor}

\long\def\com#1{}

\long\def\xxx#1{}

\defaultleftmargin{\parindent}{}{}{}

\def\cb{Crypto-Book\xspace}
\def\cw{Crypto-Wiki\xspace}
\def\cd{Crypto-Dissent\xspace}
\def\cdr{Crypto-Drop\xspace}

\begin{document}

\date{}

\title{\Large \bf \vspace{-0.5em}%
	\cb: Bootstrapping Privacy Preserving Online Identities\\
	from Social Networks}

\author{
	John Maheswaran, Daniel Jackowitz, David Isaac Wolinsky,
	Lining Wang, and Bryan Ford \\
	Yale University
}

\maketitle


\vspace{-5em}
\subsection*{Abstract}
Social networking sites supporting federated identities
offer a convenient and increasingly popular mechanism
for cross-site authentication.
Unfortunately, they also exacerbate many privacy and tracking risks.
We propose \cb, an anonymizing layer
enabling cross-site authentication while reducing
these risks.

\cb relies on a set of independently managed servers
that collectively assign each social network identity a public/private keypair.
Only an identity's owner learns all the private key shares, and can therefore
construct the private key,
while all participants can obtain any user's public key,
even if the corresponding private key has yet to be retrieved.
Having obtained an appropriate key set, a user can then leverage anonymous
authentication techniques such as linkable ring signatures to log into
third-party web sites while preserving privacy.

We have implemented a prototype of \cb
and demonstrate its use with three applications:
a Wiki system, an anonymous group communication system, and a whistleblower
submission system.
Our results show that for anonymity sets of size $100$,
\cb login takes $0.56$s for signature generation by the client,
$0.38$s for signature verification on the server, and requires
$5.6$KB of communication bandwidth.

\section{Introduction}
Social networks have gained widespread popularity
among users as a way to manage their online identity across the web.
While protocols like OAuth~\cite{oauthrfc} and OpenID~\cite{openid} allow users to maintain a single
set of credentials for cross-site authentication, such federated login can
leak privacy-sensitive profile information~\cite{golijan12privacy},
making the user's online activity more easily tracked.

To protect themselves, users could forego using social network identities
altogether, or limit the content of their social profiles.
Ideally, users could leverage their social network profiles
but in a way as to prevent third-party sites from accessing sensitive information.
While anonymous authentication protocols exist,
their practicality depend on such
technologies as PGP and personal X509 certificates --
subjects most users lack the knowledge or motivation
to use effectively.


The idea of leveraging social networks
as a basis for privacy preserving identities 
was suggested in a recent workshop paper~\cite{cryptobook}. 
Building on this idea, we present \cb,
a framework enabling users to login to third party sites
anonymously or pseudonymously via existing social network identities.
\cb interposes an anonymity layer
between existing identity management systems such as social networks
and the third-party sites users may wish to log into.
\cb prevents the social network site
from learning which external sites a user accesses,
and prevents the external site from learning
which social network user accessed their site
or which {\em other} external sites the same user has accessed.

\begin{figure}[t]
\centering
\includegraphics[width=0.30\textwidth,trim=0 0 0 0,clip]{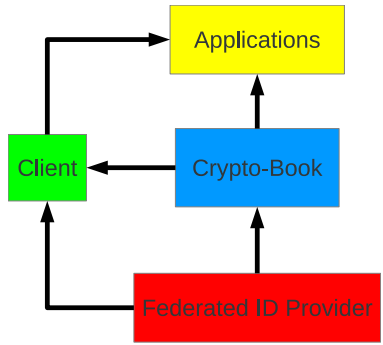}
\caption{High level system diagram}
\label{fig:high_level}
\end{figure}

Figure~\ref{fig:high_level} shows a high level overview of the architecture. The client initially authenticates with a federated ID provider. The client then collects their private key and other people's public keys from \cb which contains a set of key servers. The client generates an anonymous signature using these keys and uses that to anonymouly log in to third party applications.

Users obtain a component of their public/private key pair from each \cb key server and
use a client-side \cb module to combine these parts and
produce their composite key pair. 
The key servers have a split trust design such that \emph{all} key servers must be compromised in order to compromise a user's private key.
The key servers additionally supply a user with the
public keys corresponding to any other social networking identities.
The key servers make use of existing technologies,
such as OAuth, to verify the social network identity.

\cb performs anonymous authentication
with the acquired private and public keys
using {\em linkable ring signatures}~\cite{liu04linkable},
which prove that the signer owns {\em one} of a list of public keys,
without revealing {\em which} key.
This property is particularly useful in scenarios where
trust is associated with a group rather than an individual, such as
a journalist verifying that the source of a leaked document is a
member of a particular organization.

\cb addresses some common concerns
with online anonymity.
First, linkable ring signatures prevent a single social network identity
from performing sock-puppetry or Sybil attacks~\cite{douceur02sybil}
on websites or applications.
\cb preserves whatever Sybil attack resistance
the underlying social network offers.
Second, due to the split trust key server design,
a user need only trust that \emph{at least one} of the
key servers is honest.

We present a prototype implementation and evaluation of the \cb framework
using three applications: \cw, \cd, and \cdr.
\cw leverages \cb to provide anonymous, yet linkable
editing in a collaborative environment.
\cd shows how the integration of anonymous authentication 
with anonymous communication systems can better protect the identities
of the users of those systems.
\cdr provides anonymous document signing using \cb identities, allowing
for verifiable leaks without compromising privacy.
We built \cw using MediaWiki, the software behind Wikipedia,
and \cd using Dissent~\cite{corrigangibbs10dissent,wolinsky12scalable},
an anonymous group communication tool. \cdr extends the SecureDrop open-source
whistleblower platform.
Our experimental results show that a set of $100$ keys
requires $0.56$s for \cb signature generation on a modern laptop,
$0.38$s for signature verification by the web server,
and $5.6$KB of communication bandwidth to transmit the signature.

\cb's focus is providing convenient and usable anonymous authentication,
but it does not by itself address the general
problem of anonymous communication, especially under traffic analysis,
as systems like Tor~\cite{tor}, Dissent~\cite{dissent,dissentinnumbers}, and Aqua~\cite{aqua} do.  
\cb is synergistic with such systems but also usable independently,
when a casual level of anonymity is desired
but the user does not wish to incur the performance costs
of full anonymous forwarding.

This paper makes the following contributions:
\begin{compactitem}
  \item Privacy-preserving anonymous and pseudonymous
	authentication using social network identities.
  \item A multi-provider key distribution protocol
	preventing any single social network from impersonating a user.
  \item Experiments demonstrating the usability and practicality
	of \cb for authentication.
\end{compactitem}

\section{Background and Motivation}\label{background}

It is often difficult to strike an appropriate balance between the following objectives:

\begin{compactitem}
\item Supporting free speech and free association, and fighting censorship and oppression.
\item Improving the quality of public discourse. Hidden behind an anonymous veil, people often say or do things they might otherwise not.
\end{compactitem}

These two objectives are often at odds with each other. While Wikipedia would like to allow anonymous editing, such privileges are often abused for vandalism or sock-puppetry. We would like a system that allows users to edit pages without revealing their identity but at the same time allows the Wikipedia site administrators to sanction site abusers. We used our \cb framework to implement such an application, \cw which we describe in Section \ref{cryptowiki}.

Another example placing free speech and quality of discourse are at odds is the realm of anonymous chat. A covert organization, for example, may wish to discuss sensitive issues without revealing their individual identities while at the same time limiting access to their discusssion to people within their organization, in an effort to prevent repressive authorities or other undesirable outsiders from viewing their communications. In response, we built \cd, an application providing such functionality, on top of our \cb framework. We describe the application in detail in Section \ref{cryptodissent}.

Whistleblowing provides another such scenario, as a journalist taking possession of sensitive documents may wish to authenticate the documents without compromising the anonymity of the source. \cdr accommodates both parties by allowing documents to be signed as an anonymous member of a larger group that is credible as a whole. \cdr is described in detail in Section~\ref{cryptodrop}.

\subsection{Federated Identity}

We now give a brief overview of federated identity, focusing on OAuth~\cite{oauthrfc,oauth2}.
A federated identity protocol allows a user to present credentials
from an identity provider with which they have an account,
such as Google+/Facebook/PayPal,
and authenticate themselves to a third-party website
without revealing their actual login information (password).
Many protocols, such as OAuth, also enable the third-party to gain
limited access to the user's resources stored by the identity provider.
OpenID~\cite{openid,openid2} is another widely used federated identity protocol.

A typical OAuth login proceeds as follows,
supposing a Facebook user wishes to log into StackOverflow.

\begin{compactenum}
\item The user clicks \emph{Log in with Facebook} on StackOverflow.
\item StackOverflow redirects the user to Facebook where they login using their Facebook credentials.
\item The user gives permission for StackOverflow to access and/or modify their data (read contacts, post status updates and so on).
\item Facebook generates a temporary OAuth access token that corresponds to the granted permissions.
\item Facebook redirects the user back to StackOverflow, passing along the access token.
\item StackOverflow can now access the user's Facebook resources in line with the permissions granted by the user by including the
access token with each request.
\end{compactenum}

Using a federated identity protocol, a user can authenticate themselves to third party sites without having to maintain separate accounts for each third party website. This convenience brings privacy risks, however, of which \cb addresses the following: 
\begin{compactitem}
\item The authentication provider learns everywhere the user logs in using that identity.
\item Third party sites learn the user's true identity.
\item Users can be tracked across third party sites.
\end{compactitem}

\cb does not address the issue of stolen/compromised social network identities, where the attacker may get access to 
all of the user's accounts via a federated identity.
With this in mind, \cb does enable a user to build privacy-preserving identities
based on {\em multiple} social networking sites simultaneously -- for example both Facebook and Paypal accounts --
requiring an attacker to compromise accounts on both sites.
The general problem of stolen federated identities remains outside the scope of this paper, however.

\section{Architecture Overview}\label{architecture}

\begin{figure}[t]
\centering
\includegraphics[width=0.40\textwidth,trim=0 0 0 0,clip]{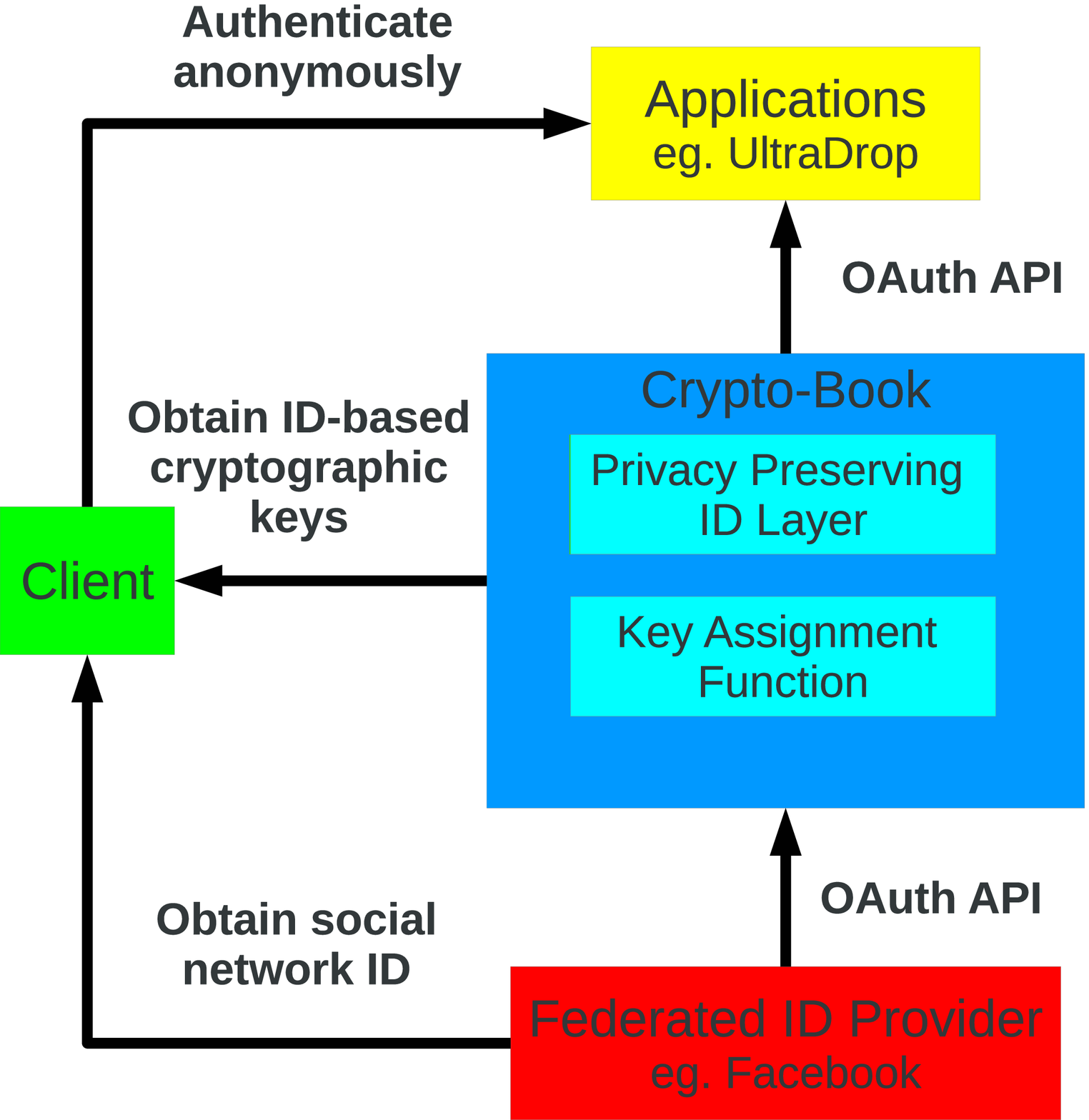}
\caption{The \cb stack}
\label{fig:stack}
\end{figure}

Figure \ref{fig:stack} shows the overall \cb stack. Traditionally, Facebook or another federated ID provider provides an API to an application via OAuth or a similar protocol. This, however, exposes the user's identity to the application. We insert the \cb layer as a buffer between the non-anonymous identity API from the ID provider and a generic, privacy preserving API for applications.

The \cb layer consists of two sub-layers:
\begin{compactenum}
\item A key assignment function
\item A privacy preserving ID layer
\end{compactenum}

The key assignment function maps public/private keypairs to each of the identities from the ID provider. The keys are then passed to the privacy-preserving ID layer, which anonymizes them to provide privacy protected IDs for the application layer through an OAuth API. The key assignment function uses linkable ring signatures~\cite{lrs} to maintain a 1-to-1 correspondence between privacy protected IDs and the original IDs, which both ensures credibility of IDs and allows applications to sanction or block abusive users -- a recourse not possible with traditional anonymous systems.

\section{Threat Model}\label{attack}
We make the following assumptions about a potential adversary in the context of our system. 
\begin{compactitem}
\item We assume that the client has the ability to connect to the key servers through an anonymity network such as Dissent~\cite{dissent,dissentinnumbers} or Tor \cite{tor}.
\item At least one of the key servers is honest; that is, it does not share its master key or private keys with anyone else. We use an $(n,n)$-threshold scheme~\cite{threshold1,threshold2,threshold3,threshold4} for the key servers (referred to in this paper as an \emph{anytrust} scheme).
\item A key server provides consistent public and private keys;  given two requests for the same key, it does not return two different results.
\item Dishonest servers may collude with each other to share master secrets or private keys.
\item Key servers can see the IP addresses of clients that connect to them. 
\item If an adversary compromises a server they have access to the master key and all private keys from that server from that epoch, but not from previous epochs (epoch details are described in Section \ref{epoch}).
\item For the \emph{single identity provider} key distribution scheme we assume that the resource server provided by the social network is honest. In this case we rely on the social network to provide identities and assume that the social network does not impersonate its users.
\item For \emph{multiple identity providers} key distribution, we assume that different identity providers (such as PayPal and Facebook) do not collude with each other to obtain a user's private key. In this case a single social network (such as Facebook) cannot compromise any user's private key.
\end{compactitem}

\section{Key Assignment}\label{keyassignment}

In this section we explain how public/private key pairs are assigned and distributed to social networking accounts. We outline at the overall server model, key distribution mechanism, key servers and compromise.

\subsection{Anytrust Server Model}\label{anytrustsec}
We use an anytrust server model in our architecture. The term anytrust was coined by Wolinsky \emph{et al.} ~\cite{anytrust} and refers to an $(n,n)$-threshold cryptosystem~\cite{threshold1,threshold2,threshold3,threshold4} where, in our case, we have $n$ key servers, and key parts from all of them are required to construct a private key.  Anytrust is a decentralized client/server
network model, in which each of many clients trust only that \emph{at least one} of a smaller but diverse set of servers behaves honestly. Moreover clients need not know which server to trust, only that at least one is honest.

We leverage anytrust in our system architecture design. Under our threat model (Section \ref{attack}), we assume only that at least one of the key servers is honest and all other key servers may be colluding to try to compromise a client's private key. For each client, each key server $i$ generates a private key part $k_i$. The client then uses some key combining function $f$ that takes each of the private key parts $k_0,k_1,\dots,k_n$ and combines them into a composite private key $k_c = f(k_0,k_1,\dots,k_n)$. The combining function $f$ is such that all $n$ key parts are required to calculate $k_c$ and any $n-1$ key parts alone provide no information about the possible value of $k_c$, thus preserving user privacy for all but $n$ key server failures.

Leveraging an anytrust server model means that we do not have to rely on a single trusted server that would be an obvious point of attack. In a single server model an adversary would have to get access to only that server's private keys to compromise a client, however in the anytrust model an adversary would have to get access to the private key parts on \emph{all} of the key servers in order to compromise the client's private key.

\subsection{Key Distribution Mechanism}\label{keydistribution}

We now explain how a client collects their private key. The primary steps for the client are as follows:
\begin{compactitem}
\item Anonymously request a link to collect their private key.
\item Authenticate with one or more social networks.
\item Collect their private key and other clients' public keys from the key servers.
\item Anonymously authenticates with third party website using keys.
\end{compactitem}

Key distribution begins with a client request for their private key. If the client simply connected to the key servers and requested their private key, this could compromise their anonymity when authenticating with third party sites as a dishonest key server could collude with a third party site to perform a timing analysis attack correlating private key pickups with subsequent authentications to the third party site. To mitigate this, the client anonymously requests a list of \cb invitations to be sent out.

Suppose Alice wants to collect her private key. She anonymously connects to a key request server (through an anonymity network) and supplies a list of social networking IDs (say Bob's and Charles'), to form an anonymity set. Each key server then sends out an invitation link to each of the social network accounts (via social networking message) to sign up and use the \cb service. This means no one knows who originally requested the invitations and hopefully several of the recipients will choose to sign up to the \cb service and collect their private key, improving anonymity and making cross site timing analysis attacks much more difficult.

Alice then receives an invitation to sign up to \cb. When she clicks the link she is redirected and has to authenticate herself with one or more social network identity providers.

\subsubsection{Alternative Identity Providers}
\cb supports alternative identity providers for example different social networks or other online identity providers such as financial institutions. \cb integrates with any federated identity provider that is OAuth compliant. Users may want to use different identity providers. For example Alice may not have a Twitter account, but has a Facebook account whereas Bob may be the other way around. Support for multiple identity providers means that more people will have access to the service. Additionally, these different identity providers may provide different \emph{identity thresholds} . For example bank-confirmed PayPal accounts have a higher barrier to entry than Facebook accounts. Using a PayPal accounts may lead to a higher degree of trust in the identities they represent.

\subsubsection{Combined Identities}\label{multipleids}
In addition to supporting alternative identity providers, \cb also allows users to authenticate with \emph{multiple} identity providers and hence obtain a private key that is tied to all of these accounts. This attests to the fact that a user has, for example, \emph{both} a Facebook and a PayPal account. Key servers may also verify that these accounts are in the same name and/or have other consistent identifying information associated with them. In the combined identities case, it is not possible for a single social network to impersonate the user and obtain their private key, as they would also need access to the user's accounts on the other identity provider sites.

Keys obtained from multiple indentity providers are combined to create a composite key as follows. Key servers use the same group parameters (for DSA based keys) or elliptic curve (for elliptic curve cryptography, EEC) and then the respective public keys can just be multiplied togther (DSA) or added (EEC) to create a combined key. The \cb client simply adds the private exponents (DSA) or scalars (EEC) together to obtain a joint private key.

\subsubsection{Private Key Pickup}
\begin{figure}[t]
\centering
\includegraphics[width=0.30\textwidth,trim=0 0 0 0,clip]{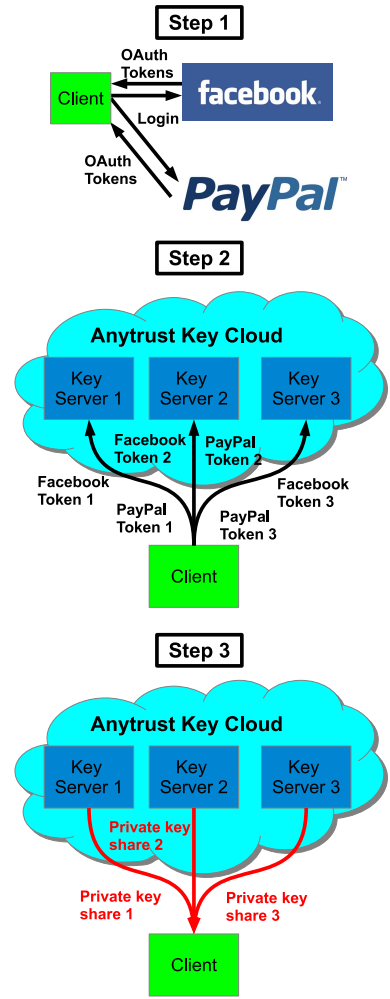}
\caption{Client authenticates with online identity providers}
\label{fig:login}
\end{figure}

Returning to our example, after Alice has received an invitation to collect her private key, she clicks the link and is taken to authenticate with one or more social networks. Upon authenticating, Alice receives one OAuth token per key server from each social network. Note that each token is only compatible with a single key server. If only one token were provided for all key servers, a malicious key server could forward the token onto other key servers to impersonate Alice. Step 1 of Figure~\ref{fig:login} shows the client authenticating with Facebook and PayPal.

Since there is one token per key server, if there are $n$ key servers, there will be $n$ corresponding Facebook (or other social network site) apps, each requiring user authorization. To simplify this process, we provide a Chrome extension that the user can download to automatically authorize all key server apps with a single Facebook signin.

Alice then forwards the appropriate OAuth tokens to the key servers (each key server receiving a token from each identity provider) as shown in step 2 of Figure \ref{fig:login}. Each key server is tasked with maintaining a public/private keypair associated with each social network identity. The key servers contact the identity providers, for example Facebook and PayPal, to verify the tokens. If verification succeeds, the key server distributes its share of the client's private key to the client as shown in step 3 of Figure \ref{fig:login}. In the case of multiple identity providers, keys are combines as described in Section \ref{multipleids}.

Once the client has received all private key shares, they combine them to get their overall composite private key using a key combining function as described in Section~\ref{anytrustsec}.

An alternative approach to having the client locally run software to collect the keys would be to have an intermediary server acting as a trusted web proxy whose job is is to authenticate the user, then collect their private key parts on their behalf, assemble them into the private key and securely return this to the client. The advantage of this is that the user does not have to run any program locally themself however they have to trust the web proxy with their private key.

\subsection{Anonymous Authentication}

\begin{figure}[t]
\centering
\includegraphics[width=0.475\textwidth,trim=0 0 0 0,clip]{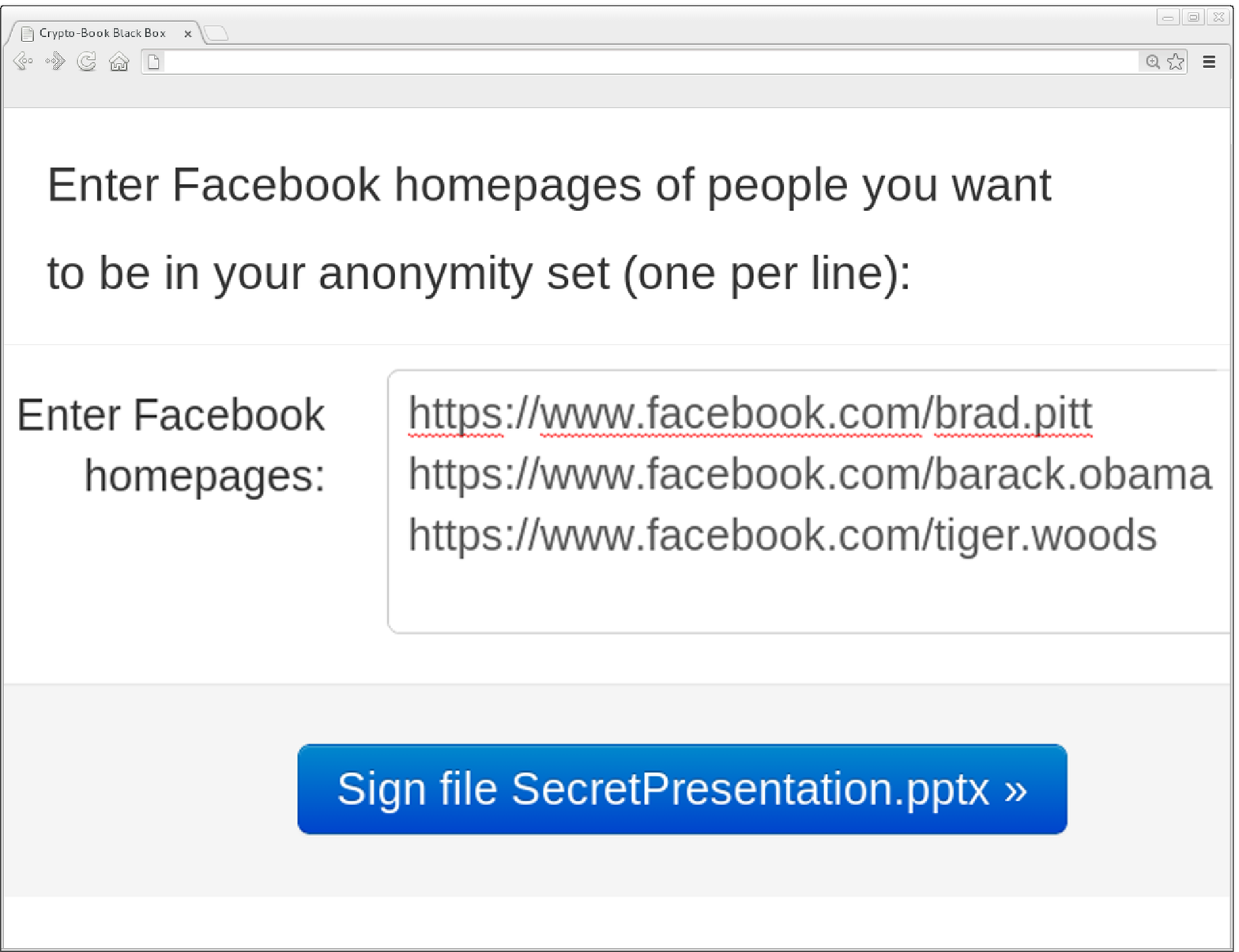}
\caption{\cb anonymity set selection}
\label{fig:crypto-sign}
\end{figure}

Once a client has obtained their own private key and a list of public keys corresponding to other Facebook identities, the client constructs a ring signature \cite{ring, ring2} with all the Facebook identities as the anonymity set. Figure \ref{fig:crypto-sign} shows our system where a user is choosing to have Brad Pitt, Barack Obama and Tiger Woods in their anonymity set by entering their Facebook profiles. A ring signature has the property that a third party can verify using only the public keys that the signature was created by one of the members of the anonymity set. However they cannot determine which person in the anonymity set specifically created the signature. Hence they can be used to protect the anonymity of the user.

The ring signature is now used by the user as a form of anonymous online identity and could be used in a multitude of different scenarios. For example the user could anonymously sign a document to give a credible leak, join an anonymous chat group open only to a specific set of users, or anonymously comment on blog posts.

\subsection{Key Servers}\label{key servers}
Each of the key servers is tasked with generating a public/private keypair for each social networking identity. There exist primarily two ways that a server could do this. The first is that a key server could generate fresh random public/private keypair on each request for either a public or private key and then store the keypair for future reference. The advantage of storing each user's keypair on disk is that is can more quickly be returned to the user as opposed to the alternative where the public or private key is calculated each time it is requested by a client.

The other way for a key server to generate public/private keypairs for clients is to hold a \emph{master secret} or \emph{master key} from which each client's private key may be created or re-created deterministically on demand to serve any given request. For example a keyed pseudo-random number generator (PRNG) such as a keyed-hash message authentication code (HMAC) could be used where the master secret is hashed together with the Facebook username of the client in order to obtain the user's private key. The advantage of this technique is that it is more scalable and requires less storage as the server does not need to store an increasing number of client keys.

\subsection{Compromised Key Servers}\label{epoch}
One threat that we need to consider is what happens if a key server's storage is compromised. For example if the master key is leaked or if a thief breaks into the data center and illicitly obtains the key server's physical hardware. The adversary would obtain access either to all the private keys saved on disk or, if the master key scheme is used, would be able to generate and hence obtain that server's part of the private key for any client. While this does not in itself compromise the user's composite private key it is still undesirable. We propose an epoch based scheme to mitigate this vulnerability.

Under this scheme we have the key server work in epochs, where the key server's master secret is valid only during a given epoch and gets randomly reinitialized in each successive epoch.  If we want previously generated ring signatures to still be verifiable after the epoch, the server must maintain a list of public keys containing at least all the public keys generated in that epoch. Then in subsequent epochs the server will be able to serve requests for older public keys (but not private keys) to allow for verification of old ring signatures.

Since the master secret gets randomly reinitialized in each successive epoch, each user can thus get a new public/private keypair in each key server epoch. The key server is only able to create (or recreate) a client's private key for the current epoch so that in case of compromise, only the user keys in the latest epoch are compromised, not those of all past epochs. We envisage epochs being realtively long, of the order of months, or new epochs triggered in the event of a key server compromise.

The epoch based key scheme used in conjunction with the anytrust key splitting scheme significantly reduces the risk of a client's private key being compromised by an adversary.

\section{Privacy Preservation}\label{privacy}

In this section we consider how user privacy is preserved in our architecture. We first look at how anonymity set choice affects user privacy, then discuss ring signatures and how they are used to provide $k$-anonymous authentication.

\subsection{Anonymity Set}\label{anonset}
One important consideration from the standpoint of maintaining user anonymity is the selection of the anonymity set used when generating the ring signature. One option is to leave it up to the user to construct their anonymity set. The benefit of this is that the user may want the set to correspond to some group of people in the real world, such as a group of employees all working in the same corporate division or government department. One a hazard of allowing users to choose their own anonymity set, however, is that they may inadvertently choose a set that compromises their anonymity. An alternative approach is for the system itself to batch users together into groups.

In some cases users may benefit from having the option of defining explicit anonymity sets. For example, a company employee may want to leak documents of public interest to the press but at the same time show that they come from a credible source. In this case the employee would like to have the option of choosing their own anonymity set where all members of the set are employees at the same company. This way, when a journalist verifies who signed the document, they know for sure that it came from an employee of the company, they just do not know which specific employee leaked the document. We implement an application that supports this functionality called \cdr that we describe in Section \ref{cryptodrop}.

The way user groups are chosen will almost certainly vary with application and may have implications for the degree of privacy protection afforded to users. The extent of these implications is an interesting area for future work which we discuss in Section \ref{nextsteps}.

A possible threat to user privacy comes from the fact that third party sites may collude to attempt to deanonymize and uniquely identify users.  If a user authenticates themself as a member of a group across many third party sites, this vector of group membership may threaten the user's anonymity. The extent of this risk depends on how the anonymity set is chosen. If different sites use different groups, for example a user is in group A on site 1 but group B on site 2, there may be some risks to user privacy the extent of which would depend on the way groups are chosen, an interesting line of investigation for future work that we mention in Section \ref{nextsteps}. 

However if groups are defined by \cb, users do not face this risk as they would always be in the same group regardless of which third party site they logged into. We propose using this scheme where \cb puts users into groups in order to protect user privacy.

\subsection{Ring signatures}
Ring signatures are a cryptographic scheme proposed by Rivest \emph{et al.} \cite{ring,ring2}. Ring signatures build on group signatures \cite{groupsig} in that a message signed with a ring signature is endorsed by someone in a particular group of people however it is difficult to determine which of the group members' keys were used to produce the signature. Ring signatures differ from group signatures in that ring signatures do not require any initial setup and can be created on an ad hoc basis. 

Rivest \emph{et al.}'s ring signatures are based on each participant having a personal RSA public private keypair. The signer must obtain the public keys of all other parties in their anonymity set, in addition to their own RSA private key in order to generate a ring signature. Rivest \emph{et al.}'s ring signatures provide the property of \emph{deniability}: Even with access to the private keys, given an RSA ring signature it is not possible to unmask the original signer. RSA keys do not support key splitting so a single key server is used as opposed to an anytrust cloud, however even if a private key is leaked it does not compromise the anonymity of existing ring signatures due to the deniability property.

Liu \emph{et al.}'s \cite{lrs} proposed \emph{linkable ring signatures} (LRS) are based on discrete logarithm DSA \cite{dsa}keys. Linkable ring signatures are similar to traditional RSA ring signatures with the additional property of  \emph{linkability}. Linkability refers to the fact that given any two signatures, a third party can determine whether or not they were produced by the same person. While LRSs provide linkability, they do not provide \emph{forward anonymity} in the way that RSA ring signatures do. If a private key is compromised by an attacker, then the attacker may use that to unmask previously generated LRSs to tell whether or not they were indeed produced by that private key.

Since DSA based LRSs do not provide deniability, we cannot trust a single server to maintain the private keys. To counteract this, a multi-server anytrust \cite{anytrust} cloud should be used to serve keys as described in Section \ref{anytrustsec}.

Linkable ring signatures are used in our architecture to provide an anonymity preserving identity to a user. The ring signature may be used to authenticate with a third party website or service so the third party knows that the user is a member of a group of users, but the specific user's identity is not revealed to the third party, protecting the user's privacy and anonymity.

\section{OAuth provider}

The \cb architecture serves as an OAuth compliant identitity provider allowing websites that include a \emph{Log in with Facebook} button to authenticate users to similarly include a \emph{Log in with \cb} button to do so anonymously and accountably. \cb as an OAuth identity provider works as follows, starting from when the third-party redirects the user to \cb to authenticate:

\cb presents a challenge to a the user in the form of a random string that the user is requested to sign using a linkable ring signature. The user signs the string and uploads their signature to \cb which then verifies the signature. If the signature is successfully verified, \cb obtains the username corresponding to that user by hashing the \emph{linkage tag} of the signature. \cb then generates an OAuth token and associates it with that username, storing the pair in a database, and redirects the user back to the third party site passing the OAuth token as a parameter.

The third party site now knows that the user successfully authenticated with \cb. The site can then include the OAuth token in future requests to query \cb for the user's username and the anonymity set they authenticated with. The fact that the username is derived from the linkage tag means that if the user tries to authenticate multiple times they will always be given the same anonymous username and thus can be held accountable for their actions on the third party site in the same way as a non-anonymous user. Hence \cb allows any third-party site or app that is an OAuth-compatible client to provide anonymous, accountable login.

\section{Implementation}\label{implementation}
To demonstrate the feasibility of our architecture, we implemented \emph{\cb}. The system allows a user to log in using Facebook, Paypal, or both, and to connect to the key servers and collect their private key. We also implemented three applications built on the \cb framework, \cw, \cd and \cdr. \cw is a Wikipedia style site where users can log in using \cb to edit the Wiki anonymously and accountably. \cd combines the \cb anonymous authentication architecture with the Dissent~\cite{dissent} chat system. \cdr is a verifiable whistleblowing application that allows users to anonymously yet credibly submit documents to journalists. We implemented OAuth provider functionality on top of \cb to allow other applications to be more easily built on top of \cb.

Our system is deployed and available online\footnote{\url{http://www.crypto-book.com}} along with source code for the system\footnote{\url{https://github.com/jyale/blackbox}}.

We used both Facebook (via Facebook Graph API~\cite{facebookgraph}) and PayPal (via PayPal Developer API~\cite{paypalapi}) as social network identity providers. We implemented DSA \cite{dsa}, RSA \cite{rsa} and Boneh-Franklin~\cite{ibe} elliptic curve based key distribution systems. As RSA does not support key splitting, our RSA implementation uses a single key server, while for the DSA and Boneh-Franklin versions we implemented a multi-server anytrust group with the keys split over the multiple servers (details in Sections \ref{dsascheme} and \ref{bonehfranklin}) and deployed our key distribution system across 10 key servers globally using PlanetLab~\cite{chun2003planetlab}. We also implemented a Google Chrome extension to automate the process of collecting a private key from the distributed servers. Since each key server is tied to a different Facebook app, the user must in turn authorize 10 Facebook apps - to automate this process we implemented a Chrome extension which requires only that the user log in to Facebook a single time in order to collect their private key.

\subsection{Key Distribution Mechanism}

We implemented \emph{two alternative ways} of collecting and assembling the key parts into a composite key:
\begin{compactitem}
\item A downloadable application allowing the user to pickup and assemble the key parts on his own machine
\item A trusted web proxy
\end{compactitem} 

Key distribution works as follows: the user logs into Facebook and PayPal and collects an OAuth token from each provider. The user then sends these tokens to each of the key servers to request their private key. Once a key server receives a private key request and corresponding OAuth token, it makes a request to the Facebook and PayPal APIs to verify that the credentials on the two accounts match and to obtain the user's corresponding Facebook username. If the authentication succeeds and a valid username is returned then the key server will lookup the corresponding private key in its database and return it to the requester (the proxy or the desktop app). For public key requests, the requester sends to the key server the Facebook username that they want to obtain the public key for, and the key server looks up the key and returns it to the requester. If for any request the server does not already have a keypair saved for that Facebook username, the server will generate a keypair and store it in its database, returning the appropriate key to the requester.

Once the requester receives responses from all of the servers it will compute the composite private and public keys. The requester, now in possession of all necessary keys, generates the linkable ring signature for the specified file.

\subsection{Key Schemes Implemented}

We implemented two different key schemes -- one based on DSA and one based on elliptic curve cryptography (Boneh-Franklin identity based encryption) which we describe below.

\subsubsection{DSA-Based Scheme}\label{dsascheme}

We implemented an LRS scheme based on DSA keys. DSA keys operate in a group $G$ of order $p$ and are of the form $Y = g^x \bmod{p}$ where $Y$ is the public key and $x$ is the private key. A composite key can be formed from a set of keys by adding the private keys and multiplying the public keys.

Our distributed key distribution relies on the fact that we can generate a composite private key $x_c$ from a list of private keys $x_0, x_1, \dots, x_n$ by summing them such that $x_c = x_0 + x_1 + \dots + x_n \bmod{q}$. The corresponding composite public key $Y_c = g^{x_0 + x_1 + \dots + x_n \bmod{q}} \bmod{p}$. This is equivalent to  multiplying the corresponding public keys $Y_c = y_0 * y_1 * \dots * y_n \bmod{p}$ so we can calculate the composite public key without knowledge of the private keys.

\subsubsection{Boneh-Franklin Identity-Based Encryption Scheme}\label{bonehfranklin}

In addition to the DSA-based scheme, we also implemented a scheme using Boneh-Franklin~\cite{ibe} identity-based encryption (IBE) keys. Boneh-Franklin is a scheme based on elliptic curves where a string such as a user's Facebook ID \emph{is} their public key. Their private key is generated for them by a private key generator (PKG). We implemented a distributed PKG where a user's private key is split among $n$ key servers using Shamir secret sharing~\cite{shamirsecret}.

A Boneh-Franklin PKG generates a user's private keys using a master private key, $s$. The PKG multiplies this by $Q_{ID}$ which is derived from the user's public key, to compute the user's private key, $Q_{priv} = sQ_{ID}$. The master private key $s$ can be split among $n$ key servers by giving each server a Shamir secret share $s_i$ of the master private key. Each key server now returns to the client a private key part $Q_{priv}^{(i)} = s_i Q_{ID}$. Using the appropriate Lagrange coeffiecients $\lambda_i$ the user can then construct their overall composite private key $Q_{priv}=\sum \lambda_{i}Q_{priv}^{(i)}$ from the private key parts $Q_{priv}^{i}$.

We demonstrated the use of Boneh-Franklin keys by implementing an encrypted Facebook messaging app that allows a user to send an encrypted Facebook message to any other Facebook user using Boneh-Franklin identity based encryption. The recipient, even if they have not previously registered with the service, can then collect their private key from the key servers and decrypt and view the message.

\section{Applications}\label{applications}

We built on our implementation from Section \ref{implementation} to develop three realistic applications using \cb. These are \cw, \cd and \cdr described below in Sections \ref{cryptowiki}, \ref{cryptodissent} and \ref{cryptodrop} respectively.

\subsection{\cw}\label{cryptowiki}
\cw is a system based on the software behind Wikipedia that allows users to log in as an \emph{accountable, anonymous} user instead of as a \emph{personally identifiable} user. This allows users to edit Wiki pages without disclosing their identity but at the same time provides resistence to abuse. Our \cw system is deployed and available online\footnote{\url{http://www.crypto-book.com}} along with full source code for the system\footnote{\url{https://github.com/jyale/cryptowiki-faction}}.

\cw provides privacy preserving accounts instead of traditional accounts. Instead of having to create a Wikipedia account in order to be able to edit pages, users instead are able to \emph{Log in with \cb} in a similar way that many sites allow users to \emph{Log in with Facebook}. When a user chooses to log in with \cb, they are redirected to the \cb website. They are then required to submit a linkable ring signature (via file upload) to the \cb servers for verification. If it is valid, the servers redirect the user back to \cw where they will log-in with an anonymized account. 

The anonymous ID is derived from the linkable ring signature that the user provides. The signatures proves that they are a member of some group and the \emph{linkage tag} is used to tell if two people signing in are the same user. If a user signs in with \cb and then abuses the Wiki site, the user's linkage tag can be blocked so they will no longer be able to access the site, ensuring accountability.

\cw site administrators are able to interact with that anonymized user in exactly the same way as with regular, non-anonymous users (which are also supported by \cw). For example administrators can message that user, interact with them on their talk page, view their edits to the Wiki and block or ban them if necessary. This allows for user accountability while simultaneously protecting user privacy.

\subsection{\cd}\label{cryptodissent}

\cd is a system that combines the \cb anonymous authentication architecture with the Dissent\cite{dissent,dissentinnumbers} chat system. Dissent is a practical group anonymity system that offers provable anonymity with accountability. \cd is an anonymous authentication platform for Dissent \cite{dissent,dissentinnumbers}, consisting of two major components:

\begin{compactitem}
  \item Anonymous Dissent chat group creation
  \item Anonymous authentication to Dissent
\end{compactitem}

\textbf{Dissent group requests} Using \cd, users request Dissent chat groups as follows: The client uses their web browser to connect to the the \cb servers. The client then generates a linkable ring signature using keys collected from the \cd key servers corresponding to the Facebook users they want in their anonymous Dissent chat group. The client sends the ring signature to the \cd servers to be verified. The \cb servers then verify the ring signature. If the signature is successfully verified by \cd, then \cd checks its database to determine whether there is already a Dissent session running for that group. If there is no active Dissent session for that group then \cd starts three Dissent servers, computes a group authentication code and inserts a corresponding record into the database along with the group members.

When a client is successfully authenticated by \cd as being a member of the group, the servers send the group authentication code to the client. The client uses this authentication code to connect to the Dissent servers: They send the authentication code to another \cd service which checks the authentication code, looks up the Dissent servers for that group and conneccts the user's Dissent client to the Dissent anonymous chat group.

\textbf{Dissent anonymous authentication} We also implemented anonymous authentication directly within Dissent using linkable ring signatures. The Dissent servers maintain a ring of public keys for that group. When a client attempts to connect to a Dissent server they are challenged to provide a valid linkable ring signature corresponding to the ring of keys for that group. The client generates this using their private key (which corresponds to one of the public keys in the ring) and sends the signature to the server. The server verifies the signature against the ring of public keys and, if successful, allows the client to connect. Our code for performing this authentication process is available on GitHub\footnote{\url{https://github.com/DeDiS/Dissent/}}.

\subsection{\cdr}\label{cryptodrop}

\begin{figure}[t]
\centering
\includegraphics[width=0.475\textwidth,trim=0 0 0 0,clip]{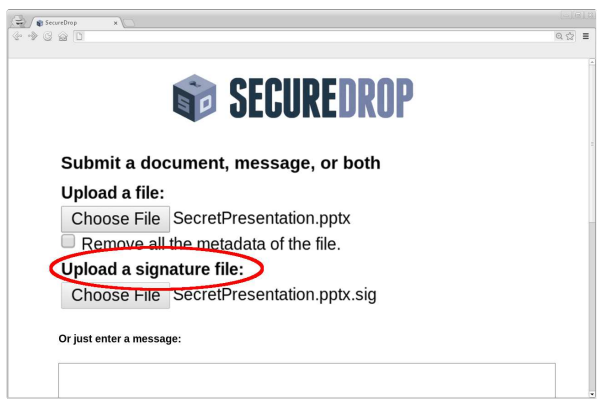}
\caption{The \cdr document source interface}
\label{fig:securedrop}
\end{figure}

\cdr builds on SecureDrop~\cite{securedrop}, an open-source whistleblower submission system managed by Freedom of the Press Foundation. SecureDrop allows journalists to accept sensitive documents from anonymous sources via a web interface. \cdr adds credibilty to leaks by allowing a source to anonymously sign a document using a relevant anonymity set and the DSA based scheme described in Section \ref{dsascheme} before submitting it via SecureDrop as shown in Figure \ref{fig:securedrop}. Upon retrieving the document, a journalist can then verify the signature, increasing confidence in the authenticity of the leak without compromising the source's anonymity.

To retrieve a verifiably leaked document, a journalist connects to the SecureDrop journalist web interface using Tor. They then download and decrypt any waiting documents using their private key. At this point, \cdr provides the additional option to verify the document by processing the signature and retrieving the signing anonymity set. This information, combined with past submissions associated with the codename, allows the journalist to make a more informed decision regarding the authenticity of the leak.

\section{Evaluation}\label{evaluation}
In order to evaluate how \cb scales with ring signature group (i.e. anonymity set) size, we carried out several experiments. We investigated the time taken for a client to generate a signature (signing phase), the time taken for the server to verify a signature (verification phase), and the size of the ring signature itself. We evaluated our \cd implementation by measuring client authentication time using our linkable ring signature scheme, also as a function of ring size. Our data suggests linear scalability in all four scenarios.

\subsection{PlanetLab deployment}

In order to evaluate the time taken for a client to collect their private key in a realistic system, we deployed key servers on a distributed set of hosts using PlanetLab~\cite{chun2003planetlab}. PlanetLab is a global research network consisting of servers located at hundreds of universities and other sites around the world. We chose 10 servers in geographically distributed locations (Texas, Florida, Illinois, France, UK, Germany, Japan, Cyprus, New Zealand, Canada) as we expect and encourage key servers to be hosted by distinct entities, preferably in different jurisdictions, in order to split trust among parties unlikely to collude and hence strengthen the anytrust model. For our experiments we deployed a single key server on each of the 10 PlanetLab nodes.

\begin{figure}[t]
\centering
\includegraphics[width=0.40\textwidth,trim=0 0 0 0,clip]{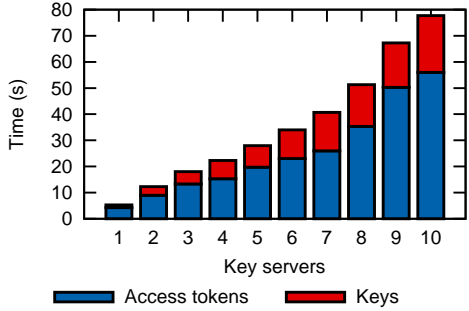}
\caption{Private key collection time}
\label{fig:key_servers}
\end{figure}

We evaluated the time required to collect a user's composite private key as a function of the number of key servers. We further divide the time into two phases: time to collect the access tokens and time to collect the key part, and present our results in Figure~\ref{fig:key_servers}. Even for a set of $10$ key servers (which is unnecessarily large due to our anytrust model), the entire key collection process completes in less than $90$ seconds. The time taken to collect the access tokens dominates as obtaining an access token requires a redirect to Facebook, execution of the Chrome extension, and redirect back; in contrast, retrieving a key is a simple web request.

\subsection{Signature generation}
\begin{figure}[t]
\centering
\includegraphics[width=0.40\textwidth,trim=0 0 0 0,clip]{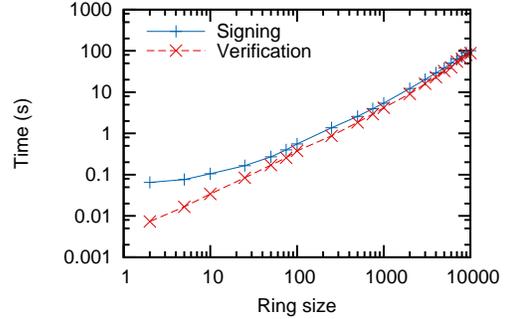}
\caption{Signing and verification time for varying ring size}
\label{fig:sign}
\end{figure}

To investigate the scalability of client-side signature generation, we ran a series of tests on our \cb code. We varied the ring size between $1000$ and $10 000$ in increments on $1000$, addressing ring sizes between $2$ and $1000$ in smaller increments. For each ring size we measured how long it took to generate a signature and present our results in Figure \ref{fig:sign}.

We observe that the time taken for signing scales roughly linearly with ring size. While a ring size of $10 000$ is far bigger than we would expect to be used in any real world application, our experimental results show good scalability even at this extreme. We expect ring sizes of at most $100$ to be used in practice, requiring on average just over half of a second ($0.56$s) for signing. 

\subsection{Verification}

We conducted a similar experiment to look at the time taken for a signature to be verified by the server. Again we considered ring sizes between $2$ and $10 000$. Or results, presented in Figure \ref{fig:sign}, indicate that the time taken for signature verification scales roughly linearly with ring size, with rings of up to $250$ memebers still taking one second on average. We expect real applications to use ring sizes of at most $100$, making verification feasible.

\subsection{Signature file size}

\begin{figure}[t]
\centering
\includegraphics[width=0.40\textwidth,trim=0 0 0 0,clip]{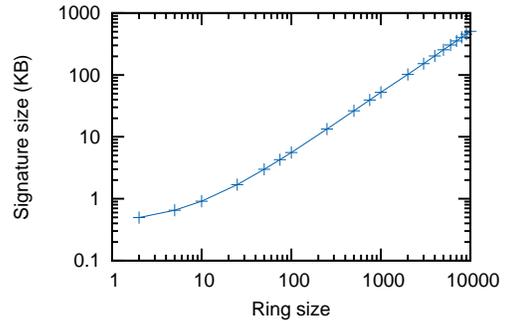}
\caption{Signature file size for varying ring size}
\label{fig:size}
\end{figure}

We also looked at how the signature file size itself varied with ring size. This is the size of the message that must be sent by the client to the server to authenticate itself. We investigated ring sizes between $2$ and $10 000$ as we did for signature generation and verification. A graph of the results is shown in Figure \ref{fig:size}. We found near-perfect linear scalability in the ring signature file size with the number of members of the ring. Ring signatures with $100$ members require only $5.6$KB which is small enough to be feasible for the client to transmit over the network to the server in a real system.

\subsection{Dissent authentication}

\begin{figure}[t]
\centering
\includegraphics[width=0.40\textwidth,trim=0 0 0 0,clip]{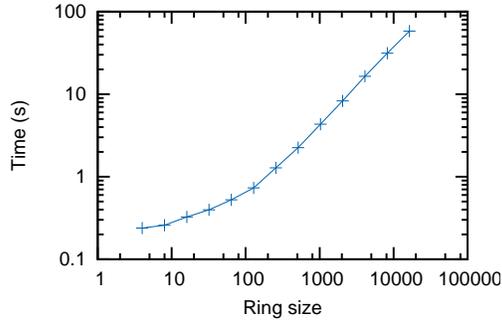}
\caption{Dissent authentication time for varying ring size}
\label{fig:dissent}
\end{figure}

Our next experiment looked at the time required for a client to authenticate itself with a Dissent server. We configured a client and server with a high capacity, low latency connection and timed at how long it took for the client to authenticate itself using our linkable ring signature authentication scheme. Authentication requires that the client generate a ring signature and send it to the server, and that the server verify the signature and notify the client. We looked at ring sizes of $2,4,8,16 \dots 16 384$ and carried out the experiment three times for each ring size. Figure \ref{fig:dissent} shows the graph of the averages. We found linear scalability in the time taken for a client to authenticate with the server.

The time taken for a client to connect to the server with no authentication was $0.01$s and the time for a client to authenticate using simple a pre-exchanged keys (DSA based public private key verification) based login was also $0.01$s. While our linkable ring signature based anonymous authentication is slower than this, it is still useable: Authentication with ring sizes up to $128$ took less than one second ($0.73$s on average for ring size $128$).

We believe that the time discrepancy between the Dissent and \cw authentication times is likely due to the fact that we implemented \cw in Python and the Dissent authentication in C++ and C++ is known to be faster than Python due to being a lower level programming language.

\subsection{Verifiable Whistleblowing}

We evaluated \cdr in the context of additional overhead in user workflow to submit a verifiable leak compared to basic, unverifiable SecureDrop. We found that, on average, the user-observable time to anonymously sign and submit a document was $66.5$ seconds, compared to $17.5$ seconds to submit an unsigned document.

From the perspective of a journalist, downloading and decrypting an unsigned document required on average $27.0$ seconds while the same process plus \cdr verification took $49.8$ seconds. Adding these time differences, the overall the \cdr user overhead for a single document was $71.8$ seconds - well within the acceptable range for practical use.

\subsection{Code modification}

We used CLOC~\cite{danial2009cloc} to count the lines of code that we had to add to modify Media Wiki, Dissent, and SecureDrop to integrate with \cb. Only small modifications were required in all cases:
\begin{compactitem}
\item {\textbf Media Wiki} required only $96$ additional lines of code to integrate it with our \cb login system (deployed externally of Media Wiki). 
\item {\textbf Dissent group request} was implemented entirely outside of the Dissent codebase, making only command line level calls to Dissent services, so required no modification of Dissent code.
\item {\textbf Dissent anonymous authentication} was implemented as an additional module to Dissent requiring $574$ additional lines of code. This is in comparison to Dissent's pre-existing authentication system that consisted of $838$ lines of code.
\item {\textbf SecureDrop} integration with our \cb signing system required only $35$ additional lines of code. 
\end{compactitem}

\section{Limitations and Future Work} \label{nextsteps}
One of the most interesting areas for future work may be in investigating the impact different anonymity set choices have on user privacy protection. In this paper we have used a scheme where users are batched into groups by \cb and where these groups are shared across all third party sites and services. Future work is required to investigate how custom, per third party service group definition can be applied without threatening user privacy.

A limitation of the current system is that linkable ring signature size scales linearly with ring size. Dodis \emph{et al.}~\cite{accring} proposed a scheme for constant space ring signatures. These signatures also have the property that both the signer and the verifier
can perform a one-time computation proportional to the size of the ring which then allows them to produce and verify
many subsequent signatures in constant time. Future work could incorporate such a scheme to reduce signing and verification time. In subsequent work, Tsang \emph{et al.}~\cite{acclrs} proposed a scheme for constant space accumulator based linkable ring signatures. Future work could incorporate such schemes to reduce signature size.

Another interesting line of inquiry may be investigating how our privacy protecting identities can be tied back into anonymous posting within Facebook as is proposed in the Faceless framework~\cite{faceless}. This would allow for anonymous discussion within existing social networking sites.

Finally, it may be worthwhile to look at what other applications could be developed on top of \cb using our privacy protecting identities. For example an anonymous group Twitter application similar to GroupTweet~\cite{grouptweet} with improved end user privacy guarantees that do not require the end user to trust the service provider with their identity.

\section{Related Work}\label{relatedwork}

The deployment of public key cryptography over social networks was considered by Narayanan \emph{et al.} \cite{socialkeys} where they considered key exchange over social networks. They considered using social networks as a public key infrastructure (PKI), they did not implement any applications that use the public keys. 

Various schemes have been proposed to protect user data \emph{within} an online social network \cite{facecloak,flybynight,noyb,safebook,easier}, by encrypting the content stored within the social network. However these schemes did not consider the privacy risks involved when a user uses their online social networking identity to identify themselves with third parties such as logging into other websites using their Facebook credentials. Dey and Weis~\cite{pseudoid} proposed PseudoID, a similar system based on blind signatures~\cite{blind} for privacy protected federated login, however their scheme does not handle key assignment or Sybil resistance as our work does. A similar blind signature based system was proposed by Khattak \emph{et al.}~\cite{khattak}. Watanabe and Miyake~\cite{pseudoid2} made initial efforts towards account checking however still did not consider key assignment. Opaak~\cite{opaak} is a system that attempts to provide some Sybil resistance through relying on a cellphone as a scare resource. SudoWeb~\cite{sudoweb} looked at limiting the amount of Facebook information disclosed to third party sites but did not consider fully anonymous online IDs.

Identity based encryption (IBE) refers to an encryption system where a public key can be an arbitrary string, for example a user's email address or social security number. The idea was first proposed by Shamir \cite{shamiribe} and since then several IBE systems have been proposed \cite{ibe1,ibe2,ibe3,ibe4,ibe5}. We take an IBE-inspired approach in assigning public keys to social network users, where the public key is deterministically computed from a user's social networking username. We also implemented a Boneh-Franklin IBE distributed PKG over Facebook.

\xxx{	Cite:
	Privacy Protection for Social Networking APIs
	http://www.cs.virginia.edu/felt/privacybyproxy.pdf
}

\section{Conclusions}\label{conclusions}
We have demonstrated \cb, a novel architecture for providing privacy preserving online identities bootstrapped off of existing social networking providers. We have implemented three major applications -- \cw, \cd, and \cdr -- on top of \cb and shown them to have good scalability properties. We believe that \cb is a usable, anonymous way to provide social network users with privacy preserving online identities. There remain a large number of areas for future research based on our architecture as well as a multitude of applications that could be developed on top of our framework leaving open a wide range of areas for investigation building on our results in future work.

\subsection*{Acknowledgments}

%
%
%
This material is based upon work supported by the Defense Advanced Research
Agency (DARPA) and SPAWAR Systems Center Pacific, Contract No. N66001-
11-C-4018.
%
%

\bibliographystyle{acm}
\bibliography{main}

\end{document}